\documentclass[twocolumn,showpacs,preprintnumbers,amsmath,amssymb]{revtex4}
\usepackage{graphicx}
\usepackage{dcolumn}
\usepackage{bm}
\begin{document}

\title{Annihilation catastrophe: From formation
to universal explosion}
\author{ Boris ~M.~Shipilevsky}
\address{ Institute of Solid State Physics, Chernogolovka,
Moscow district, 142432, Russia}
\date{\today}

\begin{abstract}
I present a systematic theory of formation of the universal
annihilation catastrophe which develops in an open system where
species $A$ and $B$ diffuse from the bulk of restricted medium and
die on its surface (desorb) by the reaction $A + B \to 0$. This
phenomenon arises in the diffusion-controlled limit as a result of
self-organizing explosive growth (drop) of the surface
concentrations of, respectively, slow and fast particles ({\it
concentration explosion}) and manifests itself in the form of an
abrupt singular jump of the desorption flux relaxation rate. As
striking results I find the dependences of time and amplitude of
the catastrophe on the initial particle number, and answer the
basic questions of when and how universality is achieved.
\end{abstract}
\maketitle

\section{INTRODUCTION}

\narrowtext For the last decades it has been shown that in spite
of the fundamental simplicity the reaction-diffusion system $A + B
\rightarrow 0$ exhibits rich cooperative behavior \cite{kot}. One
of the most impressive examples in the class of systems with {\it
bulk} reaction is the phenomenon of Ovchinnikov-Zeldovich
segregation (spontaneous growth of single species domains which
leads to {\it anomalous reaction deceleration}). Here I focus on
another wide class of systems where reaction proceeds on the {\it
catalytic surface} of medium whereas diffusion proceeds in its
bulk. In the work \cite{94} it was first demonstrated that in this
class of systems the interplay between reaction and diffusion
acquires qualitatively new features and leads to a new type of
self-organization. It has been found that once particles $A$ and
$B$ diffuse at different mobilities from the bulk of restricted
medium onto the surface and die on it (desorb) by the reaction
$A+B\rightarrow 0$, there exists some threshold difference in the
initial numbers of $A$ and $B$ particles, $\Delta_{c}$, above
which the loop of positive feedback is "switched on" and the
process of their death, instead of the usual deceleration, starts
{\it to accelerate autocatalytically}. Recently, it has been
discovered \cite{99} that the deceleration-acceleration transition
is a prelude to much more nontrivial dynamical effects : in the
diffusion-controlled limit $\Delta\to\infty$ a new critical
phenomenon develops - {\it annihilation catastrophe}, which arises
as a result of self-organizing explosive growth (drop) of the
surface concentrations of, respectively, slow and fast particles
({\it concentration explosion}) and manifests itself in the form
of an abrupt singular jump of the desorption flux relaxation rate.

The key features of the annihilation catastrophe have been
obtained on the assumption that the initial number of $A-B$ pairs
is large, so that after a transient stage the annihilation
dynamics crosses over to the independent of initial pair number
{\it universal regime} \cite{94}, \cite{99}. Until now, however,
the principal questions remain open: When and how is the universal
regime achieved? Moreover, the central question concerned with the
time moment of the catastrophe as a function of the initial pair
number remains open too. In this paper I present a systematic
theory which gives the exhaustive answers to these questions.

\section{MODEL}

I consider a model, in which species $A$ and $B$ are supposed to
be initially uniformly distributed in the bulk of an infinitely
extended slab of thickness $2\ell$. Both species diffuse to the
surface $X=\pm \ell$ $(X\in [-\ell,\ell])$ and irreversibly desorb
as a result of surface reaction $A_{ads} + B_{ads}\rightarrow AB$
with the rate proportional to the product of surface
concentrations $I=\kappa c_{As}c_{Bs}$ \cite{Sha}(Fig. 1). Because
of planar spatial homogeneity the system is effectively one
dimensional. The boundary conditions are determined from the
equality of diffusion $I^{D}$ and desorption $I$ flux densities at
the surface $I^{D}|_{s}=I$, i.e., it is assumed that the surface
layer capacity can be neglected. According to \cite{94}, \cite{99}
after introducing index $"H"$({\it heavy}) for slower diffusing
species and index $"L"$({\it light}) for a faster one, the problem
of species evolution in the dimensionless units reads (by symmetry
I consider the interval $[0,\ell]$ only)
\begin{subequations}
\label{allequations}
\begin{eqnarray}
\partial h/\partial \tau = \nabla^{2}h, \quad
\partial l/\partial \tau = (1/p)\nabla^{2}l,\label{equationa}
\\
\nabla h \mid_{s} = (1/p)\nabla l\mid_{s}= -h_{s} l_{s},\label{equationb}
\end{eqnarray}
\end{subequations}
with $\nabla (h,l) \mid_{x=0}=0$ and initial conditions $h(x ,0)=
h_{0}$ and $l(x,0)=l_{0}$. Here $h(x,\tau)=c_{H}/c_{*}$ and
$l(x,\tau)=c_{L}/c_{*}$ are the reduced concentrations, $\nabla
\equiv \partial /\partial x, x = X/\ell \in [0,1]$ is the
dimensionless coordinate, $\tau = D_{H}t/\ell^{2}$ is the
dimensionless time, $p = D_{H}/D_{L}\le 1$ is the ratio of
diffusivities, and $c_{*} = D_{H}/\kappa \ell$ is the
characteristic concentration scale. According to (1b) particles
disappear in pairs only, i.e.,
$$
J=h_{s}l_{s}=-\dot {<h>}=-\dot {<l>},
$$
where $J=I/I_{*}$ is the reduced desorption flux density
and $I_{*}=\kappa c_{*}^{2}$ is its characteristic scale,
therefore
$$
<h>-<l>=\Delta={\rm const.},
$$
i.e., the excess amount stays "inert" in the bulk (here $<h>=
\int_{0}^{1} h dx = {\cal N}_{H}/{\cal N}_{*}$ and $<l> =
\int_{0}^{1} l dx = {\cal N}_{L}/{\cal N}_{*}$ are the total
reduced numbers of particles in the bulk per unit of surface and
${\cal N}_{*}= c_{*}\ell=D_{H}/\kappa$). This "inert" part of the
majority species $\Delta=\delta{\cal N}/{\cal N}_{*}$ acts as a
control parameter, whereas its "active" part $N={\cal
N}_{pair}/{\cal N}_{*}$ (equal to the total number of $H-L$ pairs)
acts as the only variable decaying from $N_{0}$ to 0 as
$\tau\to\infty$. According to [2,3], the key features of
nontrivial dynamics, developing in the system (1) in the
diffusion-controlled limit $N_{0}=c_{L}(0)\kappa
\ell/D_{H}\to\infty$, may be formulated as follows: When parameter
$\Delta$ achieves the threshold value
\begin{eqnarray}
\Delta_{c}=\sqrt {\omega_{0}/p}\tan(\sqrt {\omega_{0}p})
\end{eqnarray}
($\omega_{0}= \pi^{2}/4$ is the main eigenfrequency of the
diffusion field relaxation), the system undergoes a transition to
the state where after a transient stage the surface concentration
of $H$-particles, $h_{s}$, and as a result the rate of death of
pairs start to grow with time autocatalytically (growth of $h_{s}$
accelerates the drop of $l_{s}$, the drop of $l_{s}$ accelerates
the growth of $h_{s}$, and so on). With growing $\Delta$ the {\it
autocatalytic stage} becomes more and more pronounced so that far
beyond the threshold self-acceleration acquires explosive
character: at $\Delta\to\infty$ the rates of growth
$\Omega_{Hs}=+\frac{d\ln h_{s}}{d\tau}$ and relaxation
$\Omega_{Ls}=-\frac{d\ln l_{s}}{d\tau}$ are synchronized
singularly growing by the law
$$
\Omega_{s}=1/|{\cal T}|, \quad |{\cal T}|=|\tau-\tau_{\star}|\to
0,
$$
where the point of {\it finite-time singularity} $\tau_{\star}$ is
achieved at the moment when the reduced number of pairs
$n(\tau)=N(\tau)/\Delta$ drops to some critical value $n_{\star}$.
The most spectacular consequence of concentration explosion is
singular behavior of the flux relaxation rate $\tau_{J}^{-1}=
-\frac {d\ln J}{d\tau}$ which is sustained constant up to the
critical point $\tau_{\star}$ upon reaching which $\tau_{J}^{-1}$
blows up abruptly to $\infty$: at ${\cal
K}=p^{3/2}\Delta/\Delta_{c}\to\infty$ the width of the jump is
contracted and its amplitude grows by the laws
$$
|{\cal T}|_{cat}\propto \Delta^{-2/5}\to 0, \quad {\rm
max}\tau^{-1}_{J}\propto \Delta^{1/4}\to\infty.
$$
In the work [3] it has been shown that in the limit $n_{0}=
N_{0}/\Delta\to\infty$ this catastrophical jump of
$\tau_{J}^{-1}$, called {\it annihilation catastrophe}, proceeds
in the {\it universal} ($n_{0}$ - independent) regime and the
scaling theory of universal explosion has been given. However, the
approach developed in [3] did not allow one to say anything
neither about the dynamics of explosion formation, nor about how
the universal regime is achieved, nor about how the point of
catastrophe depends on the initial conditions. The goal of this
paper is to give a closed theory of annihilation catastrophe
formation and, based on it, (a) to reveal the catastrophe
universalization regularities and (b) to find the dependence
$\tau_{\star}(n_{0})$ for arbitrary ratio of diffusivities. I show
that this strongly nonlinear problem allows for {\it strict and
elegant analytical solution}, I reveal its surprisingly rich
"structure", and demonstrate remarkable agreement with numerical
calculation results.

\section{Theory of universal annihilation catastrophe formation}

I start with the exact formal solution of the problem (1) in the
Laplace space $\hat{f}(s)=\hat{\cal L}f(\tau)$
\begin{eqnarray}
\nonumber
\hat{h}(x,s) = h_{0}/{s} + (\hat{h}_{s} - h_{0}/s)
\cosh(x\sqrt{s})/\cosh(\sqrt{s}),\\
\hat{l}(x,s) = l_{0}/s + (\hat{l}_{s} - l_{0}/s)
\cosh(x\sqrt{sp})/\cosh(\sqrt{sp}).
\end{eqnarray}
According to (3) the BC's (1b) acquire the form
\begin{eqnarray}
\nonumber
\hat{J} = (h_{0}/s - \hat{h}_{s})\sqrt{s}\tanh\sqrt{s} = \\
= (l_{0}/s - \hat{l}_{s})\sqrt{s/p}\tanh\sqrt{sp} = \hat{\cal L}(h_{s}l_{s})
\end{eqnarray}
and in an implicit form completely define the behavior of surface
concentrations $h_{s}$ and $l_{s}$ which in turn via Eqs.(3)
define the evolution of spatial distribution. The strategy for
solution of the nonlinear chain (4) is based on that in the
$H$-diffusion-controlled regime the $\frac{h_{s}}{<h>}$ ratio
should rapidly drop with the time, therefore, according to (4) we
can first (i) calculate $J(\tau)$ and $l_{s}(\tau)$ neglecting the
$h_{s}$ contribution, then (ii) derive $h_{s}(\tau)$ from the
condition $h_{s}= J/l_{s}$, and finally (iii) calculate
next-to-leading terms thereby defining the self-consistent picture
of surface concentrations evolution.

\subsection{Transient dynamics $\tau\ll 1$.}

At sufficiently small times the flux density is slightly changed
so assuming $J\approx J_{0}=h_{0}l_{0}$ from (4) one obtains
\begin{eqnarray}
h_{s}=h_{0}(1 - v_{h} + \cdots), \quad l_{s}=l_{0}(1 - v_{l}
+\cdots)
\end{eqnarray}
where $v_{i}=\frac {2}{\sqrt {\pi}}\sqrt {\tau/\tau_{i}}$,
$\tau_{h}=1/l_{0}^{2}$ and $\tau_{l}=1/ph_{0}^{2}$. According to
(5) relative drop rate for $l_{s}$ and $h_{s}$ is governed by the
value of the parameter
$$
R=v_{l}/v_{h}=r\sqrt {p}
$$
where $r=h_{0}/l_{0}=(1+n_{0})/n_{0}$. So, the necessary
conditions for the $H$-diffusion-controlled annihilation regime
are fulfilment of the requirements $l_{0}=N_{0}\gg 1$ and $R <
R_{c}=1$. Taking the both requirements fulfilled from (5) and (4)
one concludes that at $\tau_{h}\ll\tau\ll 1$ the flux should drop
by the law
\begin{eqnarray}
J=h_{0}(1-m)/\sqrt{\pi\tau} \approx
J_{0}(\tau_{h}/\tau)^{1/2}/\sqrt{\pi}
\end{eqnarray}
with $m\mid_{\tau/\tau_{h}\to \infty}\to 0$. In the region
$\tau_{h}\ll \tau\ll p$ where the both species diffuse in the
semi-infinite medium regime, from (4) it follows
$l_{s}-h_{s}\sqrt{p}=l_{0}\epsilon$ where $\epsilon=R_{c} - R$.
Assuming $\epsilon
> 0$ to be not too small from (4) and (6) we find
\begin{eqnarray}
h_{s}=\frac {r}{\epsilon \sqrt {\pi\tau}}(1 -\phi +\cdots), \quad
l_{s}=l_{0}\epsilon (1 + \phi +\cdots)
\end{eqnarray}
where $\phi=(R/\sqrt {\pi}\epsilon^{2})\sqrt {\tau_{h}/\tau}$.
Using then (5) and (7) we self-consistently find
$m\sim(R/\epsilon^{2})^{2}(\tau_{h}/\tau)\ll \phi$ to conclude
that asymptotics (7) is realized at the condition $\epsilon \gg
\tau^{1/4}_{h}= 1/\sqrt {N_{0}}$. In the opposite limit $0 <
\epsilon \ll \tau^{1/4}_{h}$ by the same procedure we come to the
critical asymptotics
\begin{eqnarray}
\nonumber
l_{s}(1- g + \cdots)=\sqrt {p}h_{s}(1 + g +\cdots)=\\
\nonumber \frac {l_{0}}{\pi^{1/4}}\left(\frac
{\tau_{h}}{\tau}\right)^{1/4} (1 - \phi_{c}+ \cdots),
\end{eqnarray}
where $g\sim\epsilon (\tau/\tau_{h})^{1/4}$ and
$\phi_{c}=m_{c}/2=c(\tau_{h}/\tau)^{1/4}$ with
$c=\pi^{1/4}\Gamma(\frac{3}{4})/2\Gamma(\frac{1}{4})$. At
sufficiently small $p$ and not too large $r$ (so that $R\ll
R_{c}$) in the region $\tau_{h},p \ll \tau \ll 1/r^{2}<1$ the $L$
particles distribution becomes uniform. In this regime from (4)
and (6) we find
\begin{eqnarray}
h_{s}=\frac {r}{\sqrt {\pi\tau}}(1 + \sigma +\cdots), \quad
l_{s}=l_{0}(1 - \sigma +\cdots),
\end{eqnarray}
where $\sigma = 2r\sqrt {\tau/\pi}+ O(r\sqrt {\tau_{h}}, R\sqrt
{p/\tau})$.

\subsection{Self-accelerating dynamics $\tau\gtrsim 1$.}

According to Eqs.(7), (8) at large $N_{0}\to\infty$, $\epsilon\gg
1/\sqrt{N_{0}}$ and not too large $r$ (i.e not too small $n_{0}$)
by the moment $\tau \sim 1$ when the diffusion length of $H$
particles becomes comparable with the system's size, the ratio
$h_{s}/h_{0}\propto r/\epsilon N_{0}\to 0$. Neglecting the $h_{s}$
contribution, it can be shown (see below) that at $\tau > 1$ and
large $n_{0}$ the $h_{s}$ value has to exponentially rapidly tend
to a constant ${\cal C}$. In view of this, according to (4) we
write
$$
\hat {J}=(h_{0}-{\cal C})\tanh(\sqrt {s})/\sqrt {s}
$$
whence it follows
\begin{eqnarray}
J={\cal A}e^{-\omega_{0}\tau}(1 + e^{-8\omega_{0}\tau} +\cdots),
\end{eqnarray}
where ${\cal A}=2(h_{0}-{\cal C})\approx 2h_{0}$. With the same
accuracy from (4) we have
$$
\hat {l}_{s}=l_{0}/s - [(h_{0}-{\cal C})/s]\sqrt {p}\tanh (\sqrt
{s}) \coth (\sqrt {sp})
$$
whence it follows
\begin{eqnarray}
l_{s}=({\cal A}/\Delta_{c})e^{-\omega_{0}\tau}(1 - \Lambda),
\end{eqnarray}
where $\Delta_{c}$ is defined by Eq.(2) and the leading
contribution to $\Lambda$ is governed by the sum of exponentially
decaying, $\Lambda_{-}$, and exponentially growing, $\Lambda_{+}$,
terms
$$
\Lambda = \Lambda_{-} + \Lambda_{+},
$$
\begin{eqnarray}
\Lambda_{-}=\varrho_{8}e^{-8\omega_{0}\tau} +
\varrho_{p}e^{-\chi\omega_{0}\tau}, \quad \Lambda_{+}= \lambda
e^{\omega_{0}\tau}
\end{eqnarray}
with exponent $\chi=(4/p-1)$ and amplitudes
$$
\varrho_{8}=-(\Delta_{c}/3)\sqrt {p/\omega_{0}} \cot (3\sqrt
{\omega_{0}p}),
$$
$$
\varrho_{p}=(\Delta_{c}\sqrt {p}/\pi)\tan(\pi/\sqrt {p}),
$$
$$
\lambda=\Delta_{c}(\Delta-{\cal C})/{\cal A}.
$$
From the condition $h_{s}=J/l_{s}$ and Eqs.(9),(10) locking the
chain we find
\begin{eqnarray}
h_{s}=\Delta_{c}(1+ e^{-8\omega_{0}\tau}+\cdots)/(1-\Lambda).
\end{eqnarray}
According to (11),(12) in the limit of large $\frac {\cal
A}{|\Delta -{\cal C}|\Delta_{c}}\to\infty$ ($|\lambda|\to 0$) the
$h_{s}$ value at {\it any} $\Delta$ exponentially rapidly achieves
universal asymptotics $h^{c}_{s}=\Delta_{c}$ whence it follows
${\cal C}=\Delta_{c}$. Essentially that at $p<p_{c}=4/9$ in
$\Lambda_{-}$ the first term dominates, therefore the relaxation
rate is independent of $p$, whereas at $p>p_{c}$ in $\Lambda_{-}$
the second term dominates, therefore the relaxation rate decays
with growing $p$. In view of (7),(8), and (12) using the time
convolution it is easy to check that the contribution of transient
stage is reduced only to a relative shift of the amplitude
$$
\delta {\cal A}_{tr}/{\cal
A}\mid_{h_{0}\to\infty}\sim (r/\epsilon h_{0})
\int_{0}^{O(1)}e^{\omega_{0}\theta^{2}}d\theta\to 0
$$
therefore with an accuracy of vanishingly small terms we finally
have
$$
\lambda = \Delta_{c}(\Delta-\Delta_{c})/2h_{0}.
$$
Two important consequences immediately follow from Eqs. (11),(12):
(i) at $\Delta=\Delta_{c}$ the amplitude $\lambda$ reverses its
sign which is the {\it first rigorous proof} of the threshold
arising of self-acceleration for arbitrary ratio of diffusivities
$0 < p < 1$ (note that at the long-time tail of self-acceleration
$\tau\to\infty, h_{s}\to\Delta$ a strict derivation of the
threshold $\Delta_{c}$ has been given in the work \cite{97}); (ii)
for the time of self-acceleration start, $\tau_{s}$, and departure
of the starting point $h_{s}^{\rm min}$ from the critical
asymptotics $h_{s}^{c}$, $\delta_{s}= (h_{s}^{\rm min}-
\Delta_{c})/\Delta_{c}$, we find, respectively, $\tau_{s}=
[1/\omega_{0}(\alpha +1)] \ln(\alpha\tilde{\varrho}/\lambda)$ and
\begin{eqnarray}
\delta_{s}=(\alpha+1)\tilde{\varrho}^{\frac {1}{\alpha
+1}}(\lambda/ \alpha)^{\frac {\alpha}{\alpha +1}},
\end{eqnarray}
where in the $p < p_{c}$ range $\alpha = 8,
\tilde{\varrho}=1+\varrho_{8}$ whereas in the $p_{c}<p<1$ range
$\alpha=\chi, \tilde{\varrho}=\varrho_{p}$.

\subsection{Annihilation catastrophe.}

According to (4) and (12) the exponential growth $\delta
h_{s}/h_{s}=\Lambda_{+}$ leads to the exponentially growing
contribution to the flux $\delta J/J=-\beta\Lambda_{+}^{2}$ where
$\beta= \sqrt{\omega_{0}}\tanh\sqrt{\omega_{0}}/(\Delta
-\Delta_{c})$. Far beyond the threshold $\beta\propto\Delta^{-1}$
suggesting that in the limit $\Delta\to\infty$ at the initial
stage of self-acceleration the contribution to the flux is
vanishingly small. The remarkable fact to be proved below is that
at $\Delta\to\infty$ the contribution to the flux remains
vanishingly small up to the point of finite-time singularity
$\Lambda(\tau_{\star})\to 1$ where $\dot {h_{s}}/h_{s}\to\infty$.
This means that Eqs.(9)-(12) give complete description of the
explosion dynamics. Moreover, as at $\Delta\to\infty$ the
parameter
$$
\lambda=\Delta_{c}/2(1+n_{0})
$$
becomes the unique function of $n_{0}=N_{0}/\Delta$, Eqs.(9)-(12)
allow one to achieve two {\it main goals}: a) to find the time of
the catastrophe $\tau_{\star}(n_{0})$ and b) to obtain the
description of explosion evolution with growing $n_{0}$. Taking
$\Lambda(\tau_{\star})=1$ and $\lambda < 1$ from Eqs.(11) we find
\begin{eqnarray}
\tau_{\star}=\tau^{u}_{\star}(1+\delta_{\tau}),
\end{eqnarray}
where
$$
\tau^{u}_{\star}=(4/\pi^{2}) \ln [2(1+n_{0})/\Delta_{c}]$$
and
$$
\delta_{\tau}(n_{0})= -(\varrho_{8}\lambda^{8}+
\varrho_{p}\lambda^{\chi})/|\ln\lambda|.
$$
Introducing then a relative time ${\cal T}=\tau -\tau_{\star}$
from Eq.(12) we find that at any $p<1$ in the vicinity $|{\cal
T}|\ll\omega_{0}^{-1}$ an explosive growth of $h_{s}$ sets in by
the law
\begin{eqnarray}
h_{s}=(1+Q)/\mu |{\cal T}|, \quad |{\cal T}|\to 0,
\end{eqnarray}
where $\mu=\frac{\omega_{0}}{\Delta_{c}}\sim 1-p$ and
$$
Q(n_{0})=(1+9\varrho_{8})\lambda^{8}+(1+\chi)\varrho_{p}\lambda^{\chi}.
$$
According to Eq.(9) in the course of explosion the flux is
actually "frozen"
$$
J=J_{\star}[1+\omega_{0}(1+w)|{\cal
T}|+\cdots]
$$
reaching at the point of singularity the value
\begin{eqnarray}
J_{\star}=\Delta\Delta_{c}(1+G), \quad |{\cal
T}|\to 0,
\end{eqnarray}
where $w(n_{0})=8\lambda^{8}$ and
$$
G(n_{0})=(1+\varrho_{8})\lambda^{8}+\varrho_{p}\lambda^{\chi}.
$$
From Eqs.(15),(16) and the condition $h_{s}l_{s}=J_{\star}$ there
immediately follows the synchronization of the growth,
$\Omega_{Hs}=+\frac{d\ln h_{s}}{d\tau}$, and relaxation,
$\Omega_{Ls}=-\frac{d\ln l_{s}}{d\tau}$, rates which singularly
grow by the law
$$
\Omega_{s}=\Omega_{Hs}=\Omega_{Ls}=1/|{\cal T}|.
$$
Clearly the dominant contribution to the explosion-triggered
"antiflux" $J_{ex}$ occurs in the vicinity $|{\cal T}|\ll 1$ where
the diffusional response to the explosion forms in a narrow layer
$\propto\sqrt{|{\cal T}|}$ \cite{99}. Thus, considering the medium
to be a semi-infinite one and allowing for (15) we can write
\cite{Lan}
$$
J_{ex}= -\int_{-\infty}^{{\cal T}}\frac {d h_{s}}{d\theta} \frac
{d\theta}{\sqrt{\pi({\cal T}-\theta})}\sim -\frac{(1+Q)}{\mu|{\cal
T}|^{3/2}}
$$
whence there follows smallness of $|J_{ex}|/J_{\star}$ down to
$|{\cal T}|\propto \Delta^{-2/3}\to 0$. Calculating then a
singular contribution into the flux relaxation rate
\begin{eqnarray}
\tau^{-1}_{J}=-\frac{d\ln
J}{d\tau}=\omega_{0}(1+w)+[\tau^{-1}_{J}]
\end{eqnarray}
we find
\begin{eqnarray}
[\tau^{-1}_{J}]= -\dot{J}_{ex}/J_{\star}\sim \frac{(1+Q-G)}{\Delta
|{\cal T}|^{5/2}}
\end{eqnarray}
whence there follows the {\it catastrophic jump} of
$\tau^{-1}_{J}$ from $\tau^{-1}_{J}\approx\omega_{0}$ to
$\tau^{-1}_{J}\to\infty$ with the width
$$
|{\cal T}|_{cat}\propto \Delta^{-2/5}\to 0.
$$

According to \cite{99}, the culmination consequence of the
explosion in the limit ${\cal K}\sim
p^{3/2}\Delta/\Delta_{c}\to\infty$ is the {\it exact scaling
description} of passing through the point of singularity based on
two key conditions: a) requirement
$$
\sqrt{p}\ddot{h}_{s} = \ddot{l}_{s}
$$
providing for equality of diffusional responses
$\dot{J}^{H}_{ex}=\dot{J}^{L}_{ex}$ (here and in what follows
$\dot{( )}\equiv d( )/d\tau$) and b) self-consistent condition of
passing through the singularity point at a "frozen" flux
$$
h_{s}l_{s}=J_{\star}.
$$
It has to be mentioned, however, that the scaling behavior of
$\Omega_{s}$ was postulated in \cite{99} on a basis of indirect
arguments and only later the postulated scaling function was
analytically substantiated. Moreover, the space of \cite{99} being
concise, an important chain of considerations remained beyond the
analysis given there. Below, I shall give for the first time a
systematic scaling theory of the annihilation catastrophe in the
universal limit $n_{0}\to\infty$, and then, on its basis, a
complete picture of catastrophe universalization with a growth of
$n_{0}$ will be constructed.

\subsection{Scaling laws of concentration explosion and
annihilation catastrophe.}

For simplicity, I shall begin with inferring the scaling laws of
concentration explosion and annihilation catastrophe in the
universal limit $n_{0}\to\infty$. Taking $\lambda,Q,G\to 0$,
according to (9) and (10) at the explosion stage $\omega_{0}|{\cal
T}|\to 0$ we have
$$
J^{(0)}=J_{\star}(1+\omega_{0}|{\cal T}|+ \cdots)
$$
and
$$
l_{s}^{(0)}=\mu J_{\star}|{\cal T}|(1+\omega_{0}|{\cal T}|/2 +
\cdots),
$$
whence it follows
$$
h_{s}^{ex}=J^{(0)}/l_{s}^{(0)}= 1/\mu |{\cal T}|+\cdots,
$$
where the index $"(0)"$ marks the solutions which neglect the
contribution of $h_{s}= h_{s}^{ex}$ ($h_{s}^{(0)}=0$). As it has
been mentioned above, an explosive growth of $h_{s}^{ex}$ ought to
trigger an explosive growth of "antiflux" $J_{ex}$ in calculating
which the medium can be regarded to be a semi-infinite one
\begin{equation}
J_{ex}= -\int_{-\infty}^{{\cal T}}\frac {d h_{s}^{ex}}{d\theta}
\frac {d\theta}{\sqrt{\pi({\cal T}-\theta})}.
\end{equation}
(apparently, to an accuracy of vanishingly small terms the lower
limit of the integral can be directed to $-\infty$). Substituting
here $h_{s}^{ex}$, we find
$$
J_{ex}= -a_{J}/\mu|{\cal T}|^{3/2},
$$
where $a_{J}=\sqrt{\pi}/2$. So, for the total flux we have
$$
J=J^{(0)}+ J_{ex}= J_{\star}\left(1+\omega_{0}|{\cal T}|- \frac
{a_{J}}{\mu J_{\star}|{\cal T}|^{3/2}}+\cdots\right).
$$
As the diffusion fluxes of fast and slow particles must be equal
$J^{D}_{L}|_{s}=J^{D}_{H}|_{s}=J$, it is clear that against the
background of dropping $l_{s}^{(0)}$ there must arise an explosive
growth of $l_{s}^{ex}$ which must initiate exactly the same
"antiflux" $J_{ex}^{L}=J_{ex}^{H}=J_{ex}$. Assuming that at a
developed explosion stage $\Omega_{s}p \gg 1$ the dominant
contribution to $J_{ex}$ occurs at times $|{\cal T}|/p \ll 1$,
when for $L$ particles the medium can be regarded to be
semi-infinite, by full analogy with Eq. (19) we can write
\begin{equation}
J_{ex}= -\int_{-\infty}^{{\cal T}}\frac {d l_{s}^{ex}}{d\theta}
\frac {d\theta}{\sqrt{p\pi({\cal T}-\theta})}.
\end{equation}
Thus, by requiring the equality of (19) and (20) we have
$$
l_{s}^{ex}=\sqrt{p}h_{s}^{ex}= \sqrt{p}/\mu |{\cal T}|.
$$
Making use of this result we obtain
\begin{eqnarray}
l_{s}=l_{s}^{(0)}+l_{s}^{ex}= \mu J_{\star}|{\cal T}|\left(1+
\frac{\sqrt{p}}{\mu^{2}J_{\star}{\cal T}^{2}}+ \cdots\right)
\end{eqnarray}
and finally from the condition $h_{s}=J/l_{s}$ we find
\begin{eqnarray}
h_{s}=\frac{1}{\mu |{\cal
T}|}\left(1-\frac{\sqrt{p}}{\mu^{2}J_{\star}{\cal T}^{2}}(1+
f_{J})+\cdots\right),
\end{eqnarray}
where $f_{J}= a_{J}\mu\sqrt{|{\cal T}|/p}$. From Eqs. (21) and (22) it follows that in
the vicinity of some characteristic time ${\cal T}_{f}\sim p^{1/4}/\mu
\sqrt{J_{\star}}$ the explosion rate growth begins to drastically decelerate. Due to
the requirement ${\cal T}_{f}/p << 1$ necessary for the realization of the synchronous
explosion regime (20), the explosion deceleration begins long before a noticeable flux
departure from the critical one $|J_{ex}({\cal T}_{f})/J_{\star}|\ll 1$. Introducing
the parameter
$$
{\cal K}=\mu^{2} p^{3/2}J_{\star}
$$
it can easily be seen that at any finite $0<p<1$ with a growth of
$\Delta$ in the limit of large ${\cal K}\sim
p^{3/2}\Delta/\Delta_{c}\to\infty$
$$
|J_{ex}({\cal T}_{f})|/J_{\star}\sim \sqrt{{\cal T}_{f}/p}\sim
1/{\cal K}^{1/4}\to 0.
$$
and, therefore, down to ${\cal T}_{f}\to 0$ the flux remains
"frozen". One of the most important consequences of the drastic
deceleration of the explosion rate growth is the drastic
deceleration of the flux relaxation rate growth. As it will be
shown in what follows, in the limit of large ${\cal K}\to\infty$
as a result of such deceleration the flux remains "frozen" and,
therefore, the explosion develops synchronously
\begin{eqnarray}
\Omega_{s}=\Omega_{Hs}=\Omega_{Ls}
\end{eqnarray}
{\it both before and after} passing through the point of
singularity where $\Omega_{s}$ reaches a maximum. I shall now show
that the condition (23) along with the following from (19) and
(20) {\it key condition}
\begin{eqnarray}
\dot {l}_{s}^{ex}=\sqrt{p}\dot {h}_{s}^{ex}
\end{eqnarray}
lead to a {\it remarkably complete} scaling description of passing
through the point of singularity. Taking $l_{s}=l_{s}^{(0)}+
l_{s}^{ex}$ we have
\begin{eqnarray}
\dot {l}_{s}^{ex}= \dot {l}_{s}+ {\cal C}_{(0)},
\end{eqnarray}
where ${\cal C}_{(0)}= -\dot {l}_{s}^{(0)}= \mu J_{\star}$.
Substituting (25) into (24) and using then the equalities $\dot
{h}_{s}= \Omega_{s} h_{s}$ and $\dot {l}_{s}= - \Omega_{s} l_{s}$,
following from (23), we obtain
\begin{eqnarray}
\Omega_{s}(l_{s}+ \sqrt{p}h_{s})= {\cal C}_{(0)}.
\end{eqnarray}
Differentiating (26) we find
\begin{eqnarray} \dot {\Omega}_{s}=
\Omega_{s}^{2}(l_{s}- \sqrt{p}h_{s})/(l_{s}+ \sqrt{p}h_{s}),
\end{eqnarray}
whence it follows that the explosion rate goes through the maximum
$\Omega_{s}^{M}$ in the point where
\begin{eqnarray}
l_{s}^{M}/h_{s}^{M}= \sqrt{p}.
\end{eqnarray}
Combining of (26) and (28) with the "frozen" flux condition
\begin{eqnarray}
h_{s}^{M}l_{s}^{M} = h_{s}l_{s} = J_{\star}
\end{eqnarray}
enables one to lock the chain and to readily derive the {\it
scaling law of concentration explosion}. Indeed, from (28) and
(29) we find
\begin{eqnarray}
h_{s}^{M}= p^{-1/4}\sqrt{J_{\star}}, \quad l_{s}^{M}=
p^{1/4}\sqrt{J_{\star}}.
\end{eqnarray}
Introducing then the scaling function
$$
\zeta= h_{s}/h_{s}^{M}= l_{s}^{M}/l_{s}
$$
from (26) and (30) we obtain
\begin{eqnarray}
\Omega_{s}= \frac{\dot{\zeta}}{\zeta}= \Omega_{s}^{M}F(\zeta),
\quad F(\zeta)= \frac{2\zeta}{1+ \zeta^{2}},
\end{eqnarray}
where
\begin{eqnarray}
\Omega_{s}^{M}= {\cal C}_{(0)}/2l_{s}^{M}= (\mu/2)h_{s}^{M}.
\end{eqnarray}
Integrating (31) with the proviso that $\zeta({\cal T}=0)=1$ we
easily find
\begin{eqnarray}
\zeta -1/\zeta = 2\Omega_{s}^{M}{\cal T}.
\end{eqnarray}
From Eq. (33) it follows that the characteristic time scale of
explosion is determined by the quantity ${\cal
T}_{f}=1/\Omega_{s}^{M}$ therefore introducing the reduced time
${\sf T}= {\cal T}/{\cal T}_{f}$, we finally obtain
\begin{eqnarray}
\zeta({\sf T})= {\sf T} + \sqrt{1+ {\sf T}^{2}}
\end{eqnarray}
whence it follows
\begin{eqnarray}
\Omega_{s}= \Omega_{s}^{M}S({\sf T}), \quad S({\sf T})=
\frac{1}{\sqrt{1+{\sf T}^{2}}}.
\end{eqnarray}
Two striking features of this result are symmetrical
universalization $|{\cal T}|^{-1}\leftrightarrow {\cal T}^{-1}$ of
$\Omega_{s}$ beyond the scope of interval $[-{\cal T}_{f},{\cal
T}_{f}]$ and remarkable symmetry
$$
{\cal T}\leftrightarrow -{\cal T}, \quad \zeta \leftrightarrow
1/\zeta.
$$
Substituting (34) into (19) and using (35) we come to the scaling
law of growth of the explosion-triggered "antiflux" $J_{ex}$
\begin{eqnarray}
J_{ex}= J_{ex}^{M}{\cal J}({\sf T})
\end{eqnarray}
where the amplitude at the point of explosion maximum
$$
J_{ex}^{M}= -a_{M}h_{s}^{M}\sqrt{\Omega_{s}^{M}}
$$
and the scaling function
$$
{\cal J}({\sf T})= a_{\cal J}\int_{0}^{\infty}d\theta \zeta({\sf
T}-\theta)S({\sf T}-\theta)/\sqrt{\theta}.
$$
has the asymptotics
$$
{\cal J}({\sf T})= (\pi a_{\cal J}/4)|{\sf T}|^{-3/2}, \quad -{\sf
T}\gg 1
$$
$$
{\cal J}({\sf T})= 4a_{\cal J}{\sf T}^{1/2}, \quad {\sf T}\gg 1.
$$
The coefficients $a_{M}$ and $a_{\cal J}$ are bound by the
relation $a_{M}a_{\cal J}=1/\sqrt{\pi}$ therefore by satisfying
${\cal J}(0)=1$ we find $a_{M}=\frac{2\Gamma^{2}(3/4)}{\pi}\approx
0.956$ and $a_{\cal J}= \frac{\sqrt{\pi}}{2\Gamma^{2}(3/4)}\approx
0.590$.

Differentiating (36) we find the singular part of the flux
relaxation rate in the form
\begin{eqnarray}
[\tau_{J}^{-1}]= -\dot{J}_{ex}/J_{\star}=
[\tau_{J}^{-1}]_{M}W({\sf T})
\end{eqnarray}
where the amplitude at the point of explosion maximum
$$
[\tau^{-1}_{J}]_{M}=
c_{M}h^{M}_{s}(\Omega^{M}_{s})^{3/2}/J_{\star}
$$
and the scaling function
$$
W({\sf T})= c_{W}\int_{0}^{\infty}d\theta/\sqrt {\theta}[1+({\sf
T}-\theta)^{2}]^{3/2}
$$
has the asymptotics
$$
W({\sf T})= (3\pi c_{W}/8)|{\sf T}|^{-5/2}, \quad -{\sf T}\gg 1
$$
$$
W({\sf T})= 2c_{W}{\sf T}^{-1/2}, \quad {\sf T}\gg 1.
$$
The coefficients $c_{M}$ and $c_{W}$ are bound by the relation
$c_{M}c_{W}= 1/\sqrt{\pi}$ therefore by satisfying $W(0)=1$ we
find $c_{M}= \frac{\Gamma^{2}(1/4)}{4\pi}\approx 1.046$ and
$c_{W}= \frac{4\sqrt{\pi}}{\Gamma^{2}(1/4)}\approx 0.539$. The
numerical analysis shows that at ${\sf T} = 0.46205$ the scaling
function $W({\sf T})$ reaches the maximum
$$
{\rm max} W({\sf T}) = 1.15627...
$$
whence for the amplitude of catastrophe at the point of maximum we
find
\begin{eqnarray}
 {\rm max}[\tau^{-1}_{J}]=
(1.15627...)[\tau_{J}^{-1}]_{M}.
\end{eqnarray}

Eqs. (30), (32), (34) - (37) give a detailed picture of the
concentration explosion and annihilation catastrophe in the
asymptotic limit ${\cal K}\to\infty$. Substituting into these
expressions $J_{\star}=\Delta\Delta_{c}$ and marking the
asymptotic values of amplitudes with the index $(a)$ in full
agreement with \cite{99} we obtain
\begin{eqnarray}
h_{s}^{M}(a)= p^{-1/4}\sqrt{\Delta\Delta_{c}}, \quad l_{s}^{M}(a)=
p^{1/4}\sqrt{\Delta\Delta_{c}},
\end{eqnarray}
\begin{eqnarray}
\Omega_{s}^{M}(a)= (\mu/2)p^{-1/4}\sqrt{\Delta\Delta_{c}},
\end{eqnarray}
\begin{eqnarray}
[\tau_{J}^{-1}]_{M}(a)=
(1.43340...)p^{-5/8}\Delta_{c}^{-5/4}\Delta^{1/4},
\end{eqnarray}
whence adopting (38)
$$
{\rm max}[\tau_{J}^{-1}](a)=
(1.65739...)p^{-5/8}\Delta_{c}^{-5/4}\Delta^{1/4}.
$$
From (17), (18), (37), and (41) it follows that the {\it width} of
flux relaxation rate jump does not depend on $p$
$$
|{\cal T}|_{cat}\propto \Delta^{-2/5},
$$
whereas its {\it amplitude}
$$
{\rm max}\tau_{J}^{-1}\sim [\tau_{J}^{-1}]_{M}\propto
p^{-5/8}(1-p)^{5/4}\Delta^{1/4}
$$
grows rapidly with diminishing $p$. So, {\it the less is $p$},
i.e. the less $L$-diffusion restrains the explosion development,
{\it the more brightly the effect is displayed} \cite{99}.

According to (20), one of the necessary conditions for the scaling
description of passing through the point of singularity is the
requirement
$$
\Omega_{s}p \gg 1
$$
which implies that for the medium to be considered as
semi-infinite in the process of explosion, the explosion rate must
be much beyond the characteristic rate of the $L$ particles
diffusion. Combining this requirement with Eqs. (34) and (35) and
using the equality $\Omega_{s}^{M}(a)=\sqrt{\cal K}/2p$ one can
easily see that applicability limits of the scaling description
are described by the inequalities
$$
1/\sqrt{\cal K}\ll \zeta \ll \sqrt{\cal K}, \quad  |{\sf T}| \ll
\sqrt{\cal K}.
$$
A more rigorous requirement to the quantity ${\cal K}$ is imposed
by the apparent chain of conditions
$$
\frac{|J_{ex}^{M}|}{J_{\star}}\sim
\frac{\Omega_{Ls}^{M}-\Omega_{Hs}^{M}}{\Omega_{Hs}^{M}}\sim
\frac{[\tau_{J}^{-1}]_{M}}{\Omega_{s}^{M}}\sim \mu/{\cal
K}^{1/4}\ll 1,
$$
whence it is to be expected that with a growth of ${\cal K}$ the
amplitudes (39) - (41) ought to be reached mainly by the law
$\propto \mu/{\cal K}^{1/4}$. One of the remarkable analytical
advantages of the above given approach is that it enables one not
only to determine the {\it exact asymptotic amplitudes} (39)-(41)
but, also, to answer the question of {\it when and how} they are
reached. A systematic analysis of the crossover to the asymptotics
(39)-(41) is given in the Appendix. Here I shall focus on the main
results. The central conclusion of the presented analysis is that
$\Omega_{Hs}^{M}$ reaches the asymptotic limit $\Omega_{s}^{M}(a)$
much faster than $\Omega_{Ls}^{M}$, therefore {\it the point of
the explosion maximum is defined by precisely by the point of the}
$\Omega_{Hs}$ {\it maximum}. According to (A7)
$$
\Omega_{Hs}^{M}/\Omega_{s}^{M}(a)= 1- B_{\Omega}\mu/{\cal
K}^{1/4}+ O_{\Omega}({\cal K}^{-1/2}),
$$
where $B_{\Omega}\approx 0.0318$. With taking into account the
equality $\Omega_{Ls}^{M}= \Omega_{Hs}^{M}+ \omega_{0}+
[\tau_{J}^{-1}]_{M}$ and Eq.(37) it yields
$$
\Omega_{Ls}^{M}/\Omega_{s}^{M}(a)=
1+(c_{M}/\sqrt{2}-B_{\Omega})\mu/{\cal K}^{1/4}+\cdots.
$$
According to (A8)
$$
[\tau_{J}^{-1}]_{M}/[\tau_{J}^{-1}]_{M}(a)= 1+ B_{J}\mu/{\cal
K}^{1/4}+ O_{J}({\cal K}^{-1/2})
$$
where $B_{J}\approx 0.776$. From these expressions we conclude
that $\Omega_{Hs}^{M}$ always goes to its asymptotics {\it from
below} whereas $[\tau_{J}^{-1}]_{M}$ and $ \Omega_{Ls}^{M}$ always
go to their asymptotics {\it from above}. Remarkably, the
coefficient $B_{\Omega}$ appears to be {\it anomalously small} so
that the contribution of ${\cal K}^{-1/4}$ term in the case of
$\Omega_{Hs}^{M}$ becomes less than $0.01$ already at ${\cal K}>
10^{2}$. Below I shall present extensive numerical calculations in
wide ranges of $p, \Delta$ and ${\cal K}$ which demonstrate
excellent agreement between the numerical data and the analytical
predictions.

\subsection{Universalization of annihilation catastrophe.}

Apparently the scaling theory of the annihilation catastrophe,
developed in the previous section for the universal limit
$n_{0}\to\infty$, holds in the {\it general case} of finite
$n_{0}$ too. Indeed, according to (9) and (10) at finite $n_{0}$
and $\lambda < 1$ the chain (24)-(32) remains valid with the sole
difference that now
$$
J_{\star}(n_{0})= \Delta\Delta_{c}[1+G(n_{0})]
$$
and
$$
{\cal C}_{(0)}(n_{0})= \mu J_{\star}(n_{0})[1-Q(n_{0})]
$$
become the functions of $n_{0}$. We thus have the {\it complete
scheme} to lock the chain (5)-(18) and to answer the question of
{\it when and how } the universality is reached. It remains for us
to find the central characteristic of scaling, namely, the
amplitude of explosion $\Omega^{M}_{s}(n_{0})$, and then from
Eq.(37) to derive the amplitude of catastrophe
$[\tau^{-1}_{J}]_{M}(n_{0})$.

Taking $\zeta\ll 1$ from Eq.(33) we have
$$
h_{s}=(h^{M}_{s}/2\Omega^{M}_{s})/|{\cal T}|.
$$
Matching this result with Eq.(15) thereby locking the chain
(5)-(18) we obtain
$$
\Omega^{M}_{s}=\frac{\mu}{2}(1-Q)h^{M}_{s}=\frac{\mu}{2}(1-Q)p^{-1/4}\sqrt
{J_{\star}}.
$$
Using then Eq.(16) we conclude that at $\lambda < 1$ evolution of
explosion with growing $n_{0}$ {\it is completely defined} by
functions $Q(n_{0})$ and $G(n_{0})$, and find finally the laws of
universalization of amplitudes of explosion
$$
\Omega^{M}_{s}(n_{0})=\Omega^{M}_{s}(\infty)(1+\delta_{\Omega})
$$
and catastrophe
$$
[\tau^{-1}_{J}]_{M}(n_{0})=
[\tau^{-1}_{J}]_{M}(\infty)(1+\delta_{J})
$$
in the form
\begin{eqnarray}
\delta_{\Omega}= G/2-Q, \quad \delta_{J}=G/4-3Q/2.
\end{eqnarray}
Leaving aside details here, I distinguish two main consequences of
(42):

1) According to (15),(16) the drop of $\delta_{\Omega}$ and
$\delta_{J}$ with growing $n_{0}$ is {\it surprisingly rapid}:
$$
Q,G\propto n_{0}^{-8}(p<p_{c}), \quad Q,G\propto
n_{0}^{-\chi}(p>p_{c}),
$$
where $3<\chi (p) <8$. Comparing this with the relatively slow
decrease of $\delta_{s}$[Eq.(13)]
$$
\delta_{s}(p<p_{c})\propto n_{0}^{-8/9}, \quad
\delta_{s}(p>p_{c})\propto n_{0}^{-\chi/(\chi+1)}
$$
we conclude that universalization of explosion occurs {\it long
before} $h_{s}^{\rm min}$ has reached $h^{c}_{s}$;

2) According to (42) in the range $p<p_{c}$ with decreasing $p$
some critical values $\varrho^{*}_{8,i}(p^{*}_{i})$ are reached at
which $\delta_{\Omega}$ and $\delta_{J}$ reverse their sign ($-
\rightarrow +$) so that contrary to an intuitive reasoning at
$p<p^{*}_{\Omega}$ and $p<p^{*}_{J}$ the amplitudes
$\Omega^{M}_{s}$ and ${\rm max}\tau^{-1}_{J}$, respectively, {\it
drop} with growing $n_{0}$. From (15),(16) and (42) we obtain
$$
\varrho^{*}_{8,\Omega}=-1/17,\quad  p^{*}_{\Omega}=0.0609
$$
and
$$
\varrho^{*}_{8,J}=-5/53,\quad  p^{*}_{J}=0.0217.
$$
Note, that this correlates with the behavior of the function
$\delta_{\tau}$ that pass through zero ($-\rightarrow +$) at
$$
p^{*}_{\tau}=1/9.
$$

\section{Numerical calculations}

To test analytical predictions I have carried out extensive
numerical calculations of Eqs.(1). The numerical integration of
Eqs.(1) was performed by means of the {\it implicit}
discretization scheme of increased accuracy with an additional
{\it "fictitious"} node at the surface \cite{Cra}, \cite{Smi}. The
scheme allowed performing the calculations in the system with
strong difference in species diffusivities with an accuracy down
to $10^{-3}\%$. The space and time steps were changed within the
ranges $3\times 10^{-4}\div 3\times 10^{-6}$ and $10^{-4}\div
10^{-11}$, respectively, with the number of time steps being
$10^{5}\div 10^{6}$.

In Figs. 2 and 3 are shown the results for $\Delta=10^{5}$ and
$p=0.25$, giving detailed picture of formation of the universal
explosion with growing $n_{0}$. It is seen that in accord with
Eqs.(42) already at small departures from $n^{c}_{0}(R=R_{c})$ the
transient dynamics (7) terminates in explosion that {\it
remarkably rapidly} becomes universal: further growth of $n_{0}$
leads to progressing shift of the critical point
$\tau_{\star}(n_{0})$ (Figs. 2a and 3) without changing the
explosive dynamics in its vicinity but gradually universalizing
{\it the entire self-acceleration trajectory} (Fig. 2b). I
distinguish two important points which characterize the
universalization process:

i) In accord with Eqs. (35) and (37) a {\it symmetrical} "flash"
of the explosion rate $\Omega_{Hs}$ $(|{\cal
T}|^{-1}\leftrightarrow {\cal T}^{-1})$ (Fig. 3a) and an
accompanying it {\it sharply asymmetrical} jump of the flux
relaxation rate $\tau_{J}^{-1}$ $(|{\cal T}|^{-5/2}\leftrightarrow
{\cal T}^{-1/2})$ (Fig. 3b) form {\it long before} the
universalization of the corresponding amplitudes, shifting
self-similarly with growing $n_{0}$.

ii) In accord with Eqs. (13) and (17) as $n_{0}$ grows, the
starting point of catastrophe reaches the level $\omega_{0}$ (Fig.
3b) {\it long before} the starting point of self-acceleration,
$h_{s}^{\rm min}$, reaches the level $\Delta_{c}$ (Fig. 2).

In the insets in Fig. 3 are shown the dependences
$\Gamma_{\Omega}(n_{0})=
\Omega^{M}_{Hs}(n_{0})/\Omega^{M}_{Hs}(\infty)$ and
$\tau_{\star}(n_{0})$ plotted on the basis of the dependences
$\Omega_{Hs}(\tau, n_{0})$, represented on the main panel and the
analogous dependences obtained in a wide $p$ range for $\Delta =
10^{5}$ (according to the Appendix, here and in what follows the
critical point $\tau_{\star}$ is defined to be the point of
maximum of $\Omega_{Hs}$). The inset in Fig. 3a demonstrates that,
in accord with (42), as $p$ decreases, the behavior of
$\Omega^{M}_{Hs}(n_{0})$ {\it changes qualitatively}: at $p >
p^{*}_{\Omega}$ the explosion amplitude $\Omega^{M}_{Hs}(n_{0})$
grows monotonously with growing $n_{0}$, reaching
$\Omega^{M}_{Hs}(\infty)$ {\it from below} whereas at $p <
p^{*}_{\Omega}$ the explosion amplitude first goes through a
maximum and then drops with the growing $n_{0}$, reaching
$\Omega^{M}_{Hs}(\infty)$ {\it from above}. The inset in Fig. 3b
demonstrates that at all $p$ with the growing $n_{0}$ the
numerically calculated $\tau_{\star}$ values come to the function
$\tau^{u}_{\star}(p, n_{0})$ calculated according to Eq. (14).
Moreover, in full agreement with Eq. (14) at $p > p^{*}_{\tau}$
the numerical $\tau_{\star}$ values come to $\tau^{u}_{\star}$
{\it from below} whereas at $p < p^{*}_{\tau}$ the numerical
$\tau_{\star}$ values come to $\tau^{u}_{\star}$ {\it from above}.
Below I shall present the results of the detailed numerical study
of the catastrophe universalization in a wide range of $p$ and
$\Delta$ and compare them with the predictions of (13), (14), and
(42) to plot on their basis the {\it complete} $n_{0} - p$ {\it
diagram} of universalization. Before discussing these results, my
main goals will be:

1) To demonstrate numerically how with a growth of $\Delta$ in the
vicinity of the critical point $\tau_{\star}$ a singularity forms,
and to show that with a growth of $\Delta$ and $n_{0}$ the time
dependences $\Omega_{Hs}({\cal T})$ and $[\tau^{-1}_{J}]({\cal
T})$ collapse to the predicted scaling functions $S({\sf T})$(35)
and $W({\sf T})$(37).

2) Making use of Eqs. (42) for selecting the $n_{0}^{u}(p)$ region
where the contribution of $n_{0}$ can be excluded with the
required accuracy, to study numerically the behavior of the
$\Omega_{Hs}^{M}$ and $[\tau^{-1}_{J}]_{M}$ amplitudes in a wide
range of $p$ and $\Delta$, and to demonstrate that with a growth
of ${\cal K}$ they reach their asymptotic values
$\Omega_{s}^{M}(a)$ (40) and $[\tau^{-1}_{J}]_{M}(a)$ (41) in
accord with the predictions of (A7)-(A11).

In Fig. 4 are shown the numerically calculated dependences $\Omega_{Hs}(\tau)$ that
demonstrate the formation of singularity with growing $\Delta$ at $n_{0}=4$ and
$p=0.25$. An analysis of the given data suggests that as $\Delta$ grows, the critical
point of the explosion maximum $\tau_{\star}(\Delta)$ rapidly ($\propto \Delta^{-1}$)
comes to the point of singularity $\tau_{\star}(\infty)$, calculated according to Eq.
(14), so that already at $\Delta > 10^{4}$ the ratio
$\delta\tau_{\star}(\Delta)/\tau_{\star}(\infty)$ becomes less than $0.001$. In the
inset to Fig. 4 are compared the dependences $\Omega_{Hs}(\tau)$ and
$\Omega_{Ls}(\tau)$ calculated numerically at $\Delta=10^{8}$, $n_{0}=4$ and $p=0.25$.
In accord with Eq. (35), at large $\Omega_{s}$ the curves are seen to merge in
"synchronous" explosion, asymmetrically coming apart away from the critical point. In
Fig. 5a are shown the data of Fig. 4 for $\Delta= 10^{5}, 10^{6}, 10^{7}, 10^{8}$,
replotted in the coordinates $\Omega_{Hs}- {\cal T}$ where ${\cal T}=\tau -
\tau_{\star}$, $\tau_{\star}$ being the point of the explosion maximum. Here are also
represented the data of Fig. 3a for $n_{0}= 1.8, 2.4, 3.1$ $(\Delta = 10^{5},
p=0.25)$. It is seen that in full agreement with Eq. (35) i) the rate of explosion
$\Omega_{Hs}({\cal T})$ demonstrates the {\it remarkable symmetry} $-{\cal
T}\leftrightarrow {\cal T}$ and ii) beyond the $[-{\cal T}_{f}, {\cal T}_{f}]$ region,
unlimitedly contracting with a growth of $\Delta$, the explosion rate comes to the
{\it universal law} $1/|{\cal T}|$. In the concluding Fig. 5b the data of Fig. 5a are
represented in the scaling coordinates $\Omega_{Hs}/\Omega_{Hs}^{M} - {\sf T}$ where
${\sf T}= {\cal T}/{\cal T}_{f}= {\cal T}\Omega_{Hs}^{M}$. It is seen that the
numerically calculated dependences are perfectly collapse to the scaling function
$S({\sf T})$ (35).

Let us now turn to an analysis of the flux relaxation rate. Fig. 6a demonstrates the
dependences $\tau^{-1}_{J}({\cal T})$ calculated numerically for the same parameters
as in Fig. 5a (the data are shown only for $n_{0}=4$). The data analysis suggests that
in accord with Eqs. (17), (18), and (37) with growing $\Delta$ in the vicinity of the
critical point $\tau_{\star}$ forms a {\it singular jump} of $\tau^{-1}_{J}$ the width
of which contracts unlimitedly by the law $|{\cal T}_{cat}|\propto \Delta^{-2/5}$ and
the amplitude of which grows unlimitedly by the law ${\rm max}\tau^{-1}_{J}\propto
\Delta^{1/4}$ (see below). Note that in accord with Eq. (37), after the critical point
has been passed $({\cal T}\gg {\cal T}_{f})$, the relaxation rate drops by the
$\Delta$ independent law $\propto 1/\sqrt{{\cal T}p}$, reaching at times ${\cal T}\sim
p/\omega_{0}^{2}$ the $L$-diffusion-controlled limit $\omega_{0}/p$ so that in the
limit of small $p$ there arises a most dramatic consequence of the annihilation
catastrophe: {\it an abrupt, practically instantaneous} (on the scale of $\omega_{0}$)
{\it disappearance of the flux} [3]. Based on the data of Fig. 6a in accord with Eq.
(17) the time dependences of the singular part of the flux relaxation rate were
calculated $[\tau^{-1}_{J}]({\cal T})= \tau^{-1}_{J}({\cal T}) - \omega_{0}(1+w)$
which were then replotted in the scaling coordinates $[\tau^{-1}_{J}]({\sf
T})/[\tau^{-1}_{J}](0) - {\sf T}$. The results are demonstrated in Fig. 6b. It is seen
that in perfect agreement with Eq. (37) with growing $\Delta$ the numerical results
collapse to the scaling function $W({\sf T})$. For a more detailed illustration in
Figs. 6c and 6d the data of Fig. 6b are represented in double logarithmic coordinates
in a wider range of $|{\sf T}|$ separately for ${\sf T} < 0$ (Fig. 6c) and ${\sf T} >
0$ (Fig. 6d). Ibid are given the data based on the numerical calculation for $n_{0}=
1.8, 2.4, 3.1$ $(\Delta = 10^{5}, p=0.25)$ which demonstrate a collapse to the scaling
function $W({\sf T})$ with growing $n_{0}$. It is seen from Fig. 6c that at $\Delta =
10^{8}$ the range of the scaling growth regime of $[\tau^{-1}_{J}]$ reaches six orders
of magnitude. As in this case at the starting point of growth the ratio
$[\tau^{-1}_{J}]/\omega_{0}\sim 10^{-4}$, it implies that the accuracy of the
numerical calculation of $\tau^{-1}_{J}$ reaches $10^{-3}\%$. Extensive numerical
calculations in a wide range $10^{-3}< p < 1$, a part of which will be given below,
have shown that at all the investigated $p$ values with a growth of $\Delta$ (and
therefore ${\cal K}$) the obtained numerically normalized dependences
$\Omega_{Hs}({\sf T})$ and $[\tau^{-1}_{J}]({\sf T})$ collapse, respectively, to the
scaling functions $S({\sf T})$ and $W({\sf T})$. We thus conclude that the scaling
theory of catastrophe perfectly agrees with the numerical results.

Let us now come to a numerical study of the behavior of the
amplitudes $\Omega_{Hs}^{M}(p, \Delta, n_{0})$ and
$[\tau^{-1}_{J}]_{M}(p, \Delta, n_{0})$ which are the {\it central
characteristics} of the scaling regime of catastrophe. Following
the above stated program, I shall begin with the results derived
in the {\it universal limit} $n_{0}\to\infty$. The numerical
calculations were performed within $\Delta = 10^{4}\div 10^{8}$
for $p = 0.01, 0.03, 0.1, 0.25, 0.5, 0.75$. In order that the
contribution of the initial conditions be excluded, the initial
number of particles $n_{0}$ was depending on $p$ selected from the
range $n_{0}= 10\div 200$ so that in accord with Eqs. (42) this
contribution may not exceed $10^{-3}\%$. In Fig. 7a are shown the
numerically calculated dependences
$\gamma_{\Omega}=\Omega_{Hs}^{M}/\Omega_{s}^{M}(a)$ and
$\gamma_{J}=[\tau^{-1}_{J}]_{M}/[\tau^{-1}_{J}]_{M}(a)$ as
functions of ${\cal K}/\mu^{4}$. It is seen that with growing
${\cal K}$ the numerically calculated amplitudes $\Omega_{Hs}^{M}$
and $[\tau^{-1}_{J}]_{M}$ come, respectively, to the asymptotic
values $\Omega_{s}^{M}(a)$ and $[\tau^{-1}_{J}]_{M}(a)$ calculated
analytically according to Eqs. (40), (41). Remarkably, in accord
with the predictions of Eqs. (A7), (A8) i) $\gamma_{\Omega}$ comes
to 1 {\it from below} whereas $\gamma_{J}$ comes to 1 {\it from
above}; ii) $\gamma_{\Omega}$ comes to 1 much faster than
$\gamma_{J}$; iii) the law by which $\gamma_{J}$ approaches 1 in a
wide range of ${\cal K}/\mu^{4}$ is described with excellent
accuracy by the principal term of Eq. (A8). For a more detailed
illustration of iii) in Fig. 7b in double logarithmic coordinates
are presented the dependences $\epsilon_{J}= \gamma_{J}- 1$ vs.
${\cal K}/\mu^{4}$. Ibid are shown the numerically calculated
dependences $\epsilon_{\star}= (J_{\star}- J_{M})/J_{\star}$ vs.
${\cal K}/\mu^{4}$ which demonstrate the law by which $J_{M}$
approaches $J_{\star}$ with growing ${\cal K}$. It is seen that
the numerically calculated $\epsilon_{J}$ and $\epsilon_{\star}$
values at $p$ not too close to 1 perfectly fall on the analytic
dependences $B_{J}\mu/{\cal K}^{1/4}$ [Eq. (A8)] and
$B_{\star}\mu/{\cal K}^{1/4}$ [(Eq. (A4)], respectively, shown in
thick lines (note that according to (A10) at $p\to 1, \mu\sim
1-p\to 0$ the dominant term in $\epsilon_{J}$ to ${\cal K}\propto
\mu^{-4}\to\infty$ becomes $\propto 1/\sqrt{\cal K}$). In accord
with Eqs. (A7), (A9), (A11) owing to the anomalous smallness of
the coefficient $B_{\Omega}$ the value of $\epsilon_{\Omega}=
1-\gamma_{\Omega}$ ought to decrease with growing ${\cal K}$ by
the law $\propto p/\sqrt{\cal K}$ or (at small $p$) even faster
down to very low values of $\epsilon_{\Omega}\sim 10^{-3}$. To
illustrate these predictions in Fig. 7c are shown the dependences
of $\epsilon_{\Omega}$ on $\sqrt{\cal K}$ in double logarithmic
coordinates. At comparatively high $p$ the numerically calculated
$\epsilon_{\Omega}$ values are seen to fairly fall on the analytic
dependences shown in dashed lines. As $p$ decreases the effective
"rate" of $\epsilon_{\Omega}$ drop grows only insignificantly, the
$\epsilon_{\Omega}$ value itself becomes very small already at
${\cal K}\sim 10^{2}-10^{3}$, therefore the analytical description
of $\epsilon_{\Omega}$ in this region necessitates additional
terms. Summarizing, we conclude that in the universal limit the
represented theory gives an exhaustive picture of evolution of
catastrophe and explosion amplitudes.

It only remains for me to complete this section by demonstrating
the results of extensive numerical study of the regularities of
catastrophe universalization with growing $n_{0}$. I have studied
the behavior of the dependences $\tau_{\star}(n_{0}),
\Omega^{M}_{Hs}(n_{0}), [\tau^{-1}_{J}]_{M}(n_{0})$ and $h^{\rm
min}_{s}(n_{0})$ at "scanning" $n_{0}$ from $n^{c}_{0}$ to
$10^{4}$ in wide ranges of $\Delta=10^{5}\div 10^{8}$ and
$p=10^{-3}\div 0.97$. Based on the obtained data for each of the
studied $p$ and $\Delta$ values I calculated the dependences
$\delta_{\tau}(n_{0})= \tau_{\star}(n_{0})/\tau^{u}_{\star}-1,
\delta_{\Omega}(n_{0})= \Gamma_{\Omega}(n_{0})-1,
\delta_{J}(n_{0})=
[\tau^{-1}_{J}]_{M}(n_{0})/[\tau^{-1}_{J}]_{M}(\infty)-1$ and
$\delta_{s}(n_{0})= h^{\rm min}_{s}(n_{0})/\Delta_{c}-1$ which I
then compared with the analytic predictions. I have found that in
the region of small $\delta_{i}(n_{0})$($i=\tau, \Omega, J, s$)
the behavior of functions $\delta_{i}(n_{0})$ is described with
remarkable exactness by Eqs. (14), (42) and (13). Fig. 8
represents the concluding $n_{0}-p$ {\it diagram of
universalization} where are compared the positions of the
boundaries $|\delta_{i}|=0.01(i=\tau, \Omega, J)$ and
$\delta_{s}=0.1$ resultant from a great number of numerical data
(some of which are given in the inset) for $\Delta=10^{5}$ (in the
case $i=J$ for $\Delta=10^{7}$) and calculated from (42), (14) and
(12). Excellent agreement of the analytic and numerical results
({\it not shifting} with the further growing $\Delta$ [9]) needs
no comments.

\section{Discussion and conclusion}

Finite-time singularities - blowup solutions developing from a
smooth initial conditions at a particular time - provide probably
the most dramatic manifestation of strongly nonlinear effects that
can occur in nature \cite{Kad}. The formation of finite-time
singularities is observable in a wide spectrum of nonlinear
systems (Jang-Mills fields \cite{Biz}, black holes \cite{Cho},
self-gravitating Brownian particles \cite{Cha}, turbulent flows
\cite{Pel}, jet eruption \cite{Zef}, chemotaxies \cite{Ras}, and
earthquakes \cite{Sor} to name only a few) therefore the
description of scenarios of finite-time singularities development
is a fundamental problem which attracts a wide interdisciplinary
interest.

In this paper for the first time a systematic theory of formation
of the universal annihilation catastrophe from a smooth initial
distribution has been developed and extensive numerical
calculations of the regularities of the catastrophe formation in a
wide range of parameters have been presented. The main results may
be formulated as follows:

1) The exact condition of the $H$-diffusion-controlled
annihilation regime has been found;

2) The rigorous proof of the threshold arising of
self-acceleration has been given;

3) The exact expression for the time point of catastrophe has been
obtained;

4) The closed scaling theory of passing through the point of
singularity has been given and the scaling laws of concentration
explosion and annihilation catastrophe have been derived;

5) The laws of universalization of concentration explosion and
annihilation catastrophe have been found; their surprisingly rich
"structure" has been revealed;

6) A remarkable agreement has been found between the analytical
predictions and the results of numerical calculations.

Summarizing, I believe that the analysis presented may pretend to
be one of the most striking examples of detailed description of
formation of finite-time singularity. I shall distinguish here two
most bright features of annihilation catastrophe:

i) In the majority of the models which demonstrate the formation
of finite-time singularities an analytical description of the
singularity development (based on properties of {\it
self-similarity}) appears possible only to some narrow vicinity of
the critical point beyond which the solution cannot as a rule be
continued or is principally impossible.  One of the main
advantages of the here presented theory is the asymptotically
exact scaling description of {\it passing through the point of
singularity} which yields a complete dynamical picture at the {\it
both sides} of the critical point;

ii) Arising as a result of explosive growth of the "antiflux" $J_{ex}$ at the
background of slow relaxation of the diffusion-controlled flux $J^{(0)}$, the
annihilation catastrophe demonstrates a {\it peculiar singular behavior} at which two
explosive processes ($\Omega_{Hs}$ and $\Omega_{Ls}$) are developing simultaneously,
effectively "compensating" one another so that for an external observer of flux ($J$)
the explosion dynamics {\it goes unnoticed} up to the critical point $\tau_{\star}$,
in the vicinity of which "decompensation" of explosions is manifested as a {\it sudden
singular jump} of the flux relaxation rate. In the limit of small $p$ this brings
about a most radical consequence - {\it an abrupt disappearance} of the flux.

Let us discuss in conclusion the conditions and possibilities of
an experimental observation of the annihilation catastrophe. The
irreversible bimolecular reaction $A+B\rightarrow 0$ is one of the
most abundant reactions therefore it is to be expected that the
predicted phenomena can, in principle, be observed in a wide class
of physical, chemical and biological systems with a "catalytic"
interface which, because of the high energetic barrier, does not
let diffusing particles $A$ go from medium 1 to medium 2 and
diffusing particles $B$ from medium 2 to medium 1 so that the
reaction $A+B\rightarrow 0$ can occur {\it only at the interface
between the media} \cite{Sha}, \cite{sva}, \cite{cho}. Leaving
aside here discussing systems of such type, I shall focus on the
main object of the model in question, namely, {\it adsorption -
desorption} systems (Fig. 1). Until now most of the theoretical
studies on the $A_{ads} + B_{ads}\rightarrow 0$ catalytic reaction
(the Langmuir-Hinshelwood process which is also often referred to
as the {\it monomer-monomer} catalytic scheme) have been performed
under the assumption that diffusion into the bulk can be neglected
(\cite{bur} - \cite{osh} and references therein). Such an
assumption is valid in low-temperature systems with high
surface-bulk crossover barriers, i. e., in systems with the
negligibly small bulk solubility of $A$ and $B$ particles. Here I
address to the wide class of catalytic systems where the
surface-bulk crossover barriers are not too high and, therefore,
adsorption-desorption processes are always followed by a more or
less intensive diffusion of $A$ and $B$ particles into or from the
bulk where reaction between $A's$ and $B's$ is energetically {\it
forbidden} \cite{gom}. This class of catalytic systems is not only
of fundamental interest for surface science but, also, of a
considerable applied interest for describing the interaction
kinetics of gases with metals at high temperatures (\cite{gom} -
\cite{horz} and references therein). In the work \cite{shig} the
theory has been developed for the diffusion-controlled associative
desorption of like particles, $A_{ads}+A_{ads}\rightarrow
A_{2,gas}\to 0$, from dissolved state into vacuum. Adopting this
theory, a complete picture of diffusion-controlled
thermodesorption of hydrogen and nitrogen has been constructed in
good agreement with the available experimental data. The theory
reported in this work gives a systematic description of the
diffusion-controlled kinetics of associative desorption into
vacuum of unlike particles
$$
A_{ads} + B_{ads}\rightarrow AB_{gas}\to 0,
$$
which are initially uniformly dissolved in the bulk. I shall focus
here on discussing a possibility of observation of the predicted
effects for one of the {\it most important surface reactions} of
carbon monooxide $CO$ thermodesorption from metals into vacuum
$$
C_{ads}+ O_{ads} \rightarrow CO_{gas}\to 0.
$$
It is to be mentioned first that the continual description (1)
holds as long as the "diffusion length" of the explosion at the
point of maximum remains much greater than the monolayer thickness
$a$ \cite{99}, $\delta x_{M}\sim 1/\sqrt{\Omega^{M}_{s}}\gg
a/\ell$, whence there follow the limitations
$$
\Omega^{M}_{s}\ll (\ell/a)^{2}, \quad {\cal K}\ll
p^{2}(\ell/a)^{4}.
$$
Taking, for example, $\ell/a\sim 10^{3}$ and $p\sim 0.01$ we come
to the requirements $\Omega^{M}_{s}\ll 10^{6}$ and ${\cal K}\ll
10^{8}$ to see that at any value of the reaction rate constant
$\kappa$ the specimens must have macroscopic sizes in order a
considerable effect be observed. Based on the data of monograph
\cite{fro}, I shall make estimations for three refractory metals,
i.e. niobium, tantalum, and molybdenum which at elevated
temperatures dissolve carbon and oxygen in quite large amounts.
According to \cite{fro}, at temperatures of intensive
thermodesorption of $CO$ in the range from $T\sim 1600^{0}C$ to
melting point for coefficients of carbon and oxygen diffusion in
these metals we find, respectively, $D_{C}\sim (10^{-7}\div
10^{-5}) cm^{2}/s$ and $D_{O}\sim (10^{-5}\div 10^{-4}) cm^{2}/s$,
whence it follows $p=D_{C}/D_{O}\sim 10^{-2}\div 10^{-1}$.
According to the data of \cite{fro} - \cite{horz} the desorption
rate constant of $CO$ in said temperature range alters within
$\kappa\sim (10^{-23}\div 10^{-18}) cm^{4}/s$. Substituting these
values into the expression
$$
\Delta=\delta_{C}(0)\kappa\ell/D_{C}
$$
and taking $\delta_{C}(0)= c_{C}(0)- c_{O}(0)\sim 10^{20} cm^{-3}$
and $\ell\sim 0.1 cm$ we find that in said temperature range the
$\Delta$ parameter value changes within $\Delta\sim 10^{2}\div
10^{6}$. For the density of the diffusion-controlled desorption
flux of $CO$ at the critical point we find $I_{\star}\sim
D_{C}\delta_{C}(0)/\ell\sim 10^{14}\div 10^{16} particles/cm^{2}
s$. We thus conclude that in a study of isothermal desorption of
$CO$ at elevated temperatures under high vacuum the predicted {\it
sharp jump} of the flux relaxation rate can confidently be
registered experimentally with a standard measuring technique.

\begin{acknowledgments}

This research was financially supported by the RFBR through Grants
No. 05-03-33143 and No. 02-03-33122.

\end{acknowledgments}

\appendix
\section{Crossover to the scaling regime of catastrophe.}

From (40) and (41) it follows that in the limit of large ${\cal
K}\to\infty$ the ratio
$$
\frac{\Omega_{Ls}^{M}-\Omega_{Hs}^{M}}{\Omega_{Hs}^{M}}\sim
\frac{[\tau_{J}^{-1}]_{M}}{\Omega_{s}^{M}}\propto \mu/{\cal
K}^{1/4}\to 0.
$$
I shall show below that $\Omega_{Hs}^{M}$ reaches the asymptotic
limit (40) much more rapidly than $\Omega_{Ls}^{M}$, therefore it
is the point of maximum of $\Omega_{Hs}$ that defines the point of
maximum of the explosion. With allowance for the fact that
$\Omega_{Ls}/\Omega_{Hs}\to 1$ only asymptotically, at finite
${\cal K}$ instead of (26) one has to write
\begin{eqnarray}
\Omega_{Hs}[\sqrt{p}h_{s}+(\Omega_{Ls}/\Omega_{Hs})l_{s}]={\cal
C}_{(0)}.
\end{eqnarray}
whence at the point of explosion maximum $\dot{\Omega}_{Hs}=0$ we
find
\begin{eqnarray}
\frac{\sqrt{p}h_{s}^{M}}{l_{s}^{M}}=
\left(\frac{\Omega_{Ls}^{M}}{\Omega_{Hs}^{M}}\right)^{2}-
\frac{[\dot{\tau}_{J}^{-1}]_{M}}{(\Omega_{Hs}^{M})^{2}}+
\frac{\dot{\cal C}_{(0)}}{l_{s}^{M}(\Omega_{Hs}^{M})^{2}},
\end{eqnarray}
where for completeness the term with the derivative is held
$$
\dot{\cal C}_{(0)}= -\ddot{l}_{s}^{(0)}= -\omega_{0}\mu J_{\star}
$$
and it is taken into account that from the condition
$\dot{\Omega}_{Hs}=0$ it follows $\dot{\Omega}_{Ls}=
[\dot{\tau}_{J}^{-1}]$. Differentiating (37) and calculating the
arising integral, we find the leading term in
$[\dot{\tau}_{J}^{-1}]_{M}$ in the form
$$
[\dot{\tau}_{J}^{-1}]_{M}=
\gamma_{M}h_{s}^{M}(\Omega_{s}^{M})^{5/2}/J_{\star}
$$
where $\gamma_{M}= 3\Gamma(3/4)^{2}/2\pi= (3/4)a_{M}\approx
0.717$. Substituting this result into Eq. (A2), using the equality
$\Omega_{Ls}^{M}= \Omega_{Hs}^{M}+
\omega_{0}+[\tau_{J}^{-1}]_{M}$, and taking
$\Omega_{Hs}^{M}\approx \Omega_{s}^{M}$, after separating out the
leading terms we find
\begin{eqnarray}
\sqrt{p}h_{s}^{M}/l_{s}^{M}= 1 + B_{r}\mu/{\cal K}^{1/4} +
O_{r}({\cal K}^{-1/2})
\end{eqnarray}
where
$$
B_{r}= \frac{2c_{M}-\gamma_{M}}{\sqrt{2}}\approx 0.972.
$$
At the point of the explosion maximum allowing for the
contribution of $J_{ex}$ (36) instead of (29) one has to write
$$
h_{s}^{M}l_{s}^{M}= J_{M}= J_{\star}(1+J_{ex}^{M}/J_{\star}).
$$
Substituting (36) here we have
\begin{eqnarray}
h_{s}^{M}l_{s}^{M}= J_{\star}[1-B_{\star}\mu/{\cal K}^{1/4}+
O_{\star}({\cal K}^{-1/2})],
\end{eqnarray}
where
$$
B_{\star}= a_{M}/\sqrt{2}\approx 0.676.
$$
Eqs.(A3) and (A4) immediately give
\begin{eqnarray}
h_{s}^{M}/h_{s}^{M}(a) = 1+B_{h}\mu/{\cal K}^{1/4}+ O_{h}({\cal
K}^{-1/2}),
\end{eqnarray}
where
$$
h_{s}^{M}(a)= p^{-1/4}\sqrt{J_{\star}}
$$
and
$$
B_{h}=\frac{(2c_{M}-a_{M}-\gamma_{M})}{2\sqrt{2}}\approx
0.148,
$$
and
\begin{eqnarray}
l_{s}^{M}/l_{s}^{M}(a)= 1-B_{l}\mu/{\cal K}^{1/4}+ O_{l}({\cal
K}^{-1/2}),
\end{eqnarray}
where
$$
l_{s}^{M}(a)= p^{1/4}\sqrt{J_{\star}}
$$
and
$$
B_{l}=\frac{2c_{M}+a_{M}-\gamma_{M}}{2\sqrt{2}}\approx
0.824.
$$
From (A5) and (A6) it follows that $h_{s}^{M}$ always comes to its
asymptotics $h_{s}^{M}(a)$ from above whereas $l_{s}^{M}$ always
comes to its asymptotics $l_{s}^{M}(a)$ from below. Essentially,
the coefficient $B_{h}$ is much smaller than $B_{l}$ and,
therefore, with growing ${\cal K}$ the asymptotics $h_{s}^{M}(a)$
is reached much earlier than the asymptotics $l_{s}^{M}(a)$.
Substituting then (A3) into (A1) and using (A5), (A6) we obtain
\begin{eqnarray}
\Omega_{Hs}^{M}/\Omega_{s}^{M}(a)= 1- B_{\Omega}\mu/{\cal
K}^{1/4}+ O_{\Omega}({\cal K}^{-1/2}),
\end{eqnarray}
where
$$
\Omega_{s}^{M}(a)= (\mu/2)p^{-1/4}\sqrt{J_{\star}}
$$
and
$$
B_{\Omega}= \frac{c_{M}-a_{M}}{2\sqrt{2}}\approx 0.0318.
$$
From (A7) it follows that $\Omega_{Hs}^{M}$ always comes to its
asymptotics $\Omega_{s}^{M}(a)$ from below. Remarkably, the
coefficient $B_{\Omega}$ appears so small that already at ${\cal
K}> 10^{2}$ the contribution of the ${\cal K}^{-1/4}$ term becomes
less than $0.01$. Substituting now (A4), (A5), and (A7) into the
expression
$$
[\tau_{J}^{-1}]_{M}= c_{M}h_{s}^{M}(\Omega_{Hs}^{M})^{3/2}/J_{M}
$$
for the amplitude of the catastrophe at the point of explosion
maximum we find
\begin{eqnarray}
[\tau_{J}^{-1}]_{M}/[\tau_{J}^{-1}]_{M}(a)= 1+ B_{J}\mu/{\cal
K}^{1/4}+ O_{J}({\cal K}^{-1/2}),
\end{eqnarray}
where
$$
[\tau_{J}^{-1}]_{M}(a)= (0.369834...)
\mu^{3/2}p^{-5/8}J_{\star}^{1/4}
$$
and
$$
B_{J}=\frac{a_{M}+(1/8)(2c_{M}-a_{M})}{\sqrt{2}}\approx 0.776
$$
From (A8) it follows that $[\tau_{J}^{-1}]_{M}$ always comes to
its asymptotics $[\tau_{J}^{-1}]_{M}(a)$ from above and, due to
the comparatively high $B_{J}$ value, reaches its asymptotics much
slower than $\Omega_{Hs}^{M}$.

The expressions (A5), (A6), (A7), and (A8) completely define the
principal picture of the crossover to the scaling catastrophe and
explosion regime in the limit of large ${\cal K}\to\infty$. To be
complete, I shall calculate now the corrections $O_{\Omega}({\cal
K}^{-1/2})$ and $O_{J}({\cal K}^{-1/2})$ which, as one can easily
see, may be of two types $O_{i}^{p}(p/\sqrt{\cal K})$ and
$O_{i}^{\mu}(\mu^{2}/\sqrt{\cal K})$. As $\mu\sim 1-p\to 0$ at
$p\to 1$ the $O_{i}^{p}$ corrections appear essential at $p$ close
to 1. The calculations give
\begin{eqnarray}
O_{\Omega}^{p}= -(\pi^{2}/4)p/\sqrt{\cal K}\approx -2.467
p/\sqrt{\cal K}
\end{eqnarray}
and
\begin{eqnarray}
O_{J}^{p}= \frac{(4a_{M}/c_{M}-3)\pi^{2}}{8}p/\sqrt{\cal K}\approx
+ 0.809 p/\sqrt{\cal K}.
\end{eqnarray}
Due to the anomalously low value of the coefficient $B_{\Omega}$ I
shall also give the term $O_{\Omega}^{\mu}$. The calculations
yield $O_{\Omega}^{\mu}=
[B_{\Omega}(3B_{\Omega}/2-B_{h}-B_{\star}/8)-B_{\star}^{2}/128]\mu^{2}/\sqrt{\cal
K}$ whence after substituting the coefficients we find
\begin{eqnarray}
O_{\Omega}^{\mu}\approx -0.00946 \mu^{2}/\sqrt{\cal K}.
\end{eqnarray}


\begin{references}
\bibitem{kot} E. Kotomin and V. Kuzovkov,
{\it Modern Aspects of Diffusion Controlled Reactions: Cooperative
Phenomena in Bimolecular Processes} (Elsevier, Amsterdam, 1996);
D.C. Mattis and M.L. Glasser, Rev. Mod. Phys. {\bf 70}, 979
(1998); A.J.~Bray and R.A.~Blythe, Phys. Rev. Lett. {\bf 89},
150601 (2002), and references therein.
\bibitem{94} B.M.~Shipilevsky, Phys. Rev. Lett. {\bf 73}, 201 (1994).
\bibitem{99} B.M.~Shipilevsky, Phys. Rev. Lett. {\bf 82}, 4348 (1999).
\bibitem{Sha} B.~O'Shaughnessy and D.~Vavylonis, Phys. Rev. Lett.
{\bf 84}, 3193 (2000).
\bibitem{97} B.M.~Shipilevsky, J. Phys. A {\bf 30}, L471 (1997).
\bibitem{Lan} L.D. Landau and E.M. Lifshitz, {\it Fluid Mechanics}
(Pergamon Press, Oxford, 1987).
\bibitem{Cra} J. Crank, {\it The mathematics of diffusion}
(Clarendon Press, Oxford, 1975).
\bibitem{Smi} G.D.~Smith, {\it Numerical solution of partial
differential equations with exercises and worked solutions}
(Oxford University Press, London, 1965).
\bibitem{com} Acting in the spirit of Appendix it is easy to check
that at large ${\cal K}$ and small $\delta_{\Omega, J}(n_{0})$ the
ratio $[\tau^{-1}_{J}]_{M}/\Omega^{M}_{s}\propto
\epsilon_{J}(1+\delta_{J}-\delta_{\Omega})$ so not too close to
the critical points $p^{*}_{\Omega}$ and $p^{*}_{J}$ at large
$\Delta\to\infty$ the relative deviations
$|(\delta_{i}-\delta^{\infty}_{i})/\delta^{\infty}_{i}|\propto
\epsilon_{i}\to 0$ ($i=\Omega,J$) must become vanishingly small.
\bibitem{Kad} L. Kadanoff, Phys. Today {\bf 2}, 11 (1997).
\bibitem{Biz} P. Bizon and Z. Tabor, Phys. Rev. D {\bf 64}, 121701
(2001).
\bibitem{Cho} M.W. Choptuik, Phys. Rev. Lett. {\bf 70}, 9 (1999).
\bibitem{Cha} P.H. Chavanis, C. Rosier and C. Sire, Phys. Rev. E
{\bf 66}, 036105 (2002).
\bibitem{Pel} R.B. Pelz and Y. Gulak, Phys. Rev. Lett. {\bf 79}, 4998 (1997).
\bibitem{Zef} B.W. Zeff, B. Kleber, J. Fineberg and D.P. Lathrop,
Nature (London) {\bf 403}, 401 (2000).
\bibitem{Ras} M. Rascle and C. Ziti, J. Mat. Biol. {\bf 33}, 388 (1995).
\bibitem{Sor} D. Sornette and A. Helmstetter, Phys. Rev. Lett. {\bf 89}, 158501 (2002).
\bibitem{sva} B. O'Shaughnessy and D. Vavylonis, Eur. Phys. J. E {\bf 1}, 159 (2000).
\bibitem{cho} B. Chopard, M. Droz, J. Magnin, and Z. Racz, Phys. Rev. E {\bf 56},
5343 (1997).
\bibitem{bur} A.A. Ovchinnikov and S.F. Burlatsky, JETP Lett. {\bf 43}, 638 (1986).
\bibitem{ana} L.W. Anacker and R. Kopelman, Phys. Rev. Lett. {\bf 58}, 289
(1987).
\bibitem{lin} K. Lindenberg, B.J. West and R. Kopelman, Phys. Rev. Lett.
{\bf 60}, 1777 (1988).
\bibitem{doer} D. ben-Avraham and C.R. Doering, Phys. Rev. A {\bf 37}, 5007
(1988).
\bibitem{cle} E. Clement, L.M. Sander and R. Kopelman, Phys. Rev. A
{\bf 39} 6455 (1989); {\bf 39} 6466 (1989).
\bibitem{krap} P.L. Krapivsky, Phys. Rev. A {\bf 45}, 1067 (1992).
\bibitem{eva} J.W. Evans and T.R. Ray, Phys. Rev. E {\bf 47}, 1018 (1993).
\bibitem{frac} L. Frachebourg, P.L. Krapivsky and S. Redner, Phys. Rev. Lett.
{\bf 75}, 2891 (1995).
\bibitem{osh} P. Argyrakis, S.F. Burlatsky, E. Clement and G. Oshanin,
Phys. Rev. E {\bf 63}, 021110 (2001).
\bibitem{gom} R. Gomer, {\it Chemisorption on metals}, Solid State
Physics, {\bf 30} (Academic Press, New-York, 1975).
\bibitem{fro} E. Fromm and G. Gebhardt, {\it Gase and Kohlenstoff in
Metallen} (Springer, Berlin, 1976).
\bibitem{gra} H. J. Grabke and G. Horz, {\it Kinetics and mechanisms of
gas-metal interactions}, Ann. Rev. Mater. Sci. {\bf 7}, 155
(1977).
\bibitem{horz} G. Horz, {\it Refractory Metal-Gas Systems: Thermodynamics,
Kinetics, and Mechanisms}, in: Processing and Applications of
High-Purity Refractory Metals and Alloys, Eds. P. Kumar, H. A.
Jehn, and M.Uz (Minerals, Metals and Material Society, Warrendale,
1998).
\bibitem{shig} B.M. Shipilevsky and V.G. Glebovsky, Surf. Sci. {\bf
216}, 509 (1989).
\end{references}
\end{document}